# Exploring the impact of automated vehicles lane-changing behavior on urban network efficiency

Alberto Pelizza[1]

Federico Orsini[1,2]

Sefa Yilmaz-Niewerth[3]

Riccardo Rossi[1]

Bernhard Friedrich[3]

[1]Department of Civil, Environmental and Architectural Engineering, University of Padua, Italy
[2] MoBe – Mobility and Behavior Research Center, University of Padua, Italy
[3] Institute of Transportation and Urban Engineering, Technische Universität Braunschweig, Germany

# Exploring the impact of automated vehicles lane-changing behavior on urban network efficiency

Alberto Pelizza
ICEA Department
University of Padua
Padua, Italy
alberto.pelizza@studenti.unipd.it

Federico Orsini
ICEA Department
MoBe Research Center
University of Padua
Padua, Italy
federico.orsini@dicea.unipd.it

Sefa Yilmaz-Niewerth
Institute of Transportation and Urban Engineering
Technische Universität Braunschweig
Braunschweig, Germany
s.yilmaz-niewerth@tu-braunschweig.de

Riccardo Rossi
ICEA Department
MoBe Research Center
University of Padua
Padua, Italy
riccardo.rossi@unipd.it

Bernhard Friedrich
Institute of Transportation and Urban Engineering
Technische Universität Braunschweig
Braunschweig, Germany
friedrich@tu-braunschweig.de

*Abstract*—While automated vehicle (AV) research has grown steadily in recent years, the impact of automated lane-changing behavior on transportation systems remains a largely understudied topic. The present work aims to explore the effects of automated lane-changing behavior on urban network efficiency as the penetration rate of AVs increases. To the best of the authors' knowledge, this represents the first attempt to do so by isolating the effects of the lane-changing behavior; this was obtained by considering AVs with automated lateral control, yet retaining the same longitudinal control characteristics of conventional vehicles (CV). An urban road network located in Hannover, Germany, was modeled with the microsimulation software SUMO, and several scenarios were analyzed, starting from a baseline with only CVs and then progressively increasing the AV penetration rate with 10% increments. Results highlight a modest, but statistically significant, decrease in system performance, with travel times increasing, and average speed and network capacity decreasing, as penetration rates increase. This was likely caused by a more prudent behavior of AVs, which accepted larger gaps than CVs when performing lane-changing maneuvers.

*Keywords—automated vehicles, lane-changing, SUMO, microsimulation, urban network, mixed traffic, MFD*

## I. INTRODUCTION

As the diffusion of partially-automated vehicles increases, there is an ever-growing interest in the estimation of the effects that higher-level automated vehicles (AV) will have on transportation systems in terms of efficiency, safety and emissions ([1]–[5]), with somewhat contrasting results in different contexts.

Most existing studies analyzed the effect of AV diffusion by focusing only on automated longitudinal control of vehicles ([6]–[8]). Lane-changing behavior of AVs has received less interest, especially in urban scenarios. In particular, since in previous studies considering lane-changing behavior the simulated AVs were modeled both in terms of longitudinal control (with car-following models) and lateral control (with lane-changing models) ([9]–[11]), it was not possible to clearly isolate the specific impacts that AVs lane-changing behavior has on traffic.

Here, a different approach to the problem is followed, and instead of modeling both longitudinal and lateral automated control at the same time, only the former is considered. This means that the simulated AVs in the present study follow an automated lane-changing behavior, while keeping a conventional car-following behavior. While it is of course not realistic to envisage the real-world production of vehicles that are automated exclusively in terms of lateral control, it allows us to isolate the effect of lane-changing behavior and to study its impact on network efficiency, which is the ultimate goal of the present study. To achieve it, several simulations were carried out using SUMO microsimulation software ([12], [13]), progressively increasing AV penetration rate, with lane-changing parameters chosen in accordance with existing literature studies that calibrated these parameters with real-world data (see Section II).

## II. RELATED WORKS

In order to model the automated lane-changing behavior of vehicles, a literature review was carried out, focusing on SUMO-based works. This review also serves as a reference to compare the results of the present work.

Mintsis et al. [14] modeled and calibrated the car-following and lane-changing models implemented in SUMO for both automated and conventional vehicles (CVs). With reference to the lane-changing model, a sensitivity analysis was performed on four SUMO parameters, evaluating which input parameters had more influence on the safe longitudinal gap to the leading vehicle in the ego lane and to the leading and following vehicle in the target lane. From their analysis it emerged that the `lcAssertive` was the most relevant parameter, as it significantly affected all the types of gaps. Therefore, they focused exclusively on the calibration of that parameter, using data from Hyundai Motor Europe Technical Center. It was observed that, as the speed increased, the accepted gap grew linearly; furthermore, high values of `lcAssertive` indicated a more aggressive behavior as the vehicle-driver accepts smaller and smaller gaps, while low values of `lcAssertive` indicated a more conservative behavior. The parameterized model was then applied in eight different scenarios, with different simulation runs that included the simultaneous presence of automated and conventional vehicles at different penetration rates. In general, the authors observed that AVs had more prudent behavior than CVs. Mintsis et al.'s work was taken as a reference in several other publications both in terms of the methodology used and of the calibrated parameters, especially for the car-following model ([15]–[22]).

Kavas-Torris et al. [23] evaluated the interaction between automated and conventional vehicles and how average speed, travel duration, and time loss varied as the level of autonomy increased. They used different parameters for each level of automation, resulting in more frequent lane-changing maneuvers as the automation increased. The parameters were chosen with sensitivity analysis, without any direct link to real-world data. Results showed higher average speed and reduced trip duration and time loss, suggesting that AVs could potentially bring benefits to the overall transportation system.

Lu et al. [7] investigated the impact of AVs on the capacity of an urban network through the use of macroscopic fundamental diagrams. Two scenarios were modeled within SUMO microsimulation software: a virtual grid network and a real-world road system in Budapest. The authors evaluated the impact of AVs on network capacity at different penetration rates. Only the longitudinal control of AVs was modeled, while CVs were modeled using SUMO default parameters. The authors observed an increase in the capacity of the entire network by up to 16% compared to the baseline scenario, with a linear increase in capacity as the penetration rate increased. They explained this by citing shorter headway and lower reaction time of AVs.

Nippold et al. [24] evaluated the effects of AVs in interrupted flow conditions, modeling the urban network of Dusseldorf, Germany, within SUMO; successively, they investigated the effects of AVs within a portion of the freeway. The results were based on the penetration rate of AVs, with an increase of 10% capacity in each scenario. They did not specifically model any parameters related to the lane-changing model.

The aim of Berrazouane et al. [25] was to compare the effects of AV penetration rates on motorway traffic. When calibrating CV parameters, a sensitivity analysis was carried out to understand which parameters influenced the lane-changing behavior the most, observing that the `lcAssertive` parameter influenced its behavior more, with less contribution from the `lcSpeedGain` parameter. Regarding the modeling of AVs, the authors used the calibrated values from Mintsis et al. Having built the fundamental flow-density diagram for each scenario, the authors noted that AVs have a more conservative behavior than CVs, with a reduction in terms of maximum flow as the number of AVs on the road increased.

From this brief overview, it is possible to make several remarks. First, the only existing reference for real-world calibrated lane-changing parameters in SUMO is Mintsis et al [14]. Second, there is a lack of studies investigating or even considering lane-changing behavior when modeling AVs, especially when investigating urban scenarios ([7], [24]). Third, most of the studies observed an increase in network efficiency as the penetration rate increased, except for Berrazouane et al. [25].

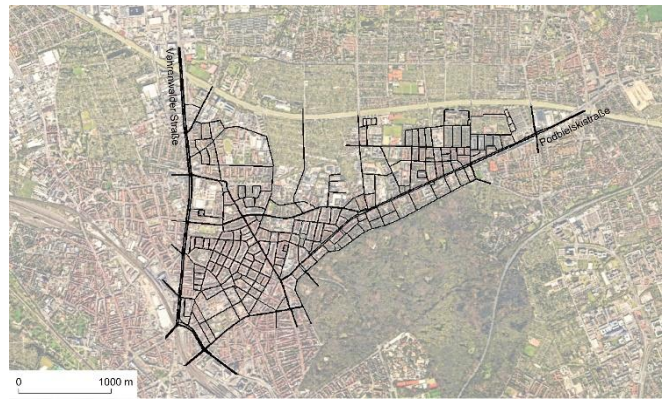

Fig. 1. Urban road network of Vahrenwald-List (Hannover, Germany)

## III. METHODOLOGY

### A. Network setup

The network analyzed was part of the road system in the Vahrenwald-List neighborhood of Hannover, Germany (Fig.1). The neighborhood is located northeast of the center of Hannover and is the most populous district of the city, with more than 70,000 inhabitants. The network covered a 5.88 km$^2$ area, containing 84 intersections (46 of them are signalized) and had a total length of 117.45 km, 39.35 of them being two-lane roads. The speed limit was 50 km/h on the main arteries and 30 km/h on local roads.

### B. Vehicle setup

As explained in Section I, in this work, AVs are modeled only in terms of lane-changing behavior, which is parametrized in SUMO with the LC2013 model [26]. LC2013 considers four types of lane-changing maneuvers: strategic, cooperative, tactical and regulatory. Strategic maneuvers are initiated to avoid a dead-end lane; cooperative maneuvers involve the collaboration of a vehicle in the target lane, which allows the ego vehicle to make the change; tactical maneuvers are performed by the ego vehicle to overtake a blocking vehicle and gain speed; regulatory maneuvers are carried out to clear the left lane, after an overtake.

SUMO AV parameters were set using as main reference the work of Mintsis et al. [14], which, as mentioned in Section II is the only lane-changing SUMO-based work calibrated with real-world data, and focused exclusively on the `lcAssertive` parameter. Since other literature studies ([23], [25]) pointed out that other parameters have a non-negligible effect, the `lcStrategic`, `lcSpeedGain`, and `lcKeepRight` were also changed from their default values, following the indications of Kavas-Torris et al. [23]. Table I shows the parameters used to describe the lane-changing behavior of CVs and AVs. Default values have been left for all the parameters that do not appear in the table.

The higher values of the AVs `lcStrategic` parameter result in an earlier lane change, meaning that compared to the CVs, they tend to change lanes in advance of the dead-end point. Higher values of the `lcSpeedGain` parameter result in more lane changes, i.e., there is a lower threshold value that causes them to change lanes more often in order to obtain a speed gain. Higher values of the `lcKeepRight` parameter correspond to a lower threshold value that makes them change lanes earlier, with the aim of leaving the fast lane free. Lower values of the `lcAssertive` parameter increase the accepted gap value in the destination lane; this produces more prudent behavior for the AVs.

TABLE I.  SUMO PARAMETERS SET FOR MODELING LANE-CHANGING BEHAVIOR OF CONVENTIONAL AND AUTOMATED VEHICLES

|  | Conventional vehicles | | | Automated vehicles | | |
|---|---|---|---|---|---|---|
| *Parameter* | *Cons.* | *Moder.* | *Aggr.* | *Cons.* | *Moder.* | *Aggr.* |
| `lcStrategic` | 1.0 | 1.0 | 1.0 | 3.0 | 3.0 | 3.0 |
| `lcSpeedGain` | 1.0 | 1.0 | 1.0 | 5.0 | 5.0 | 5.0 |
| `lcKeepright` | 1.0 | 1.0 | 1.0 | 1.2 | 1.2 | 1.2 |
| `lcAssertive` | 1.0 | 1.3 | 1.6 | 0.5 | 0.7 | 0.9 |

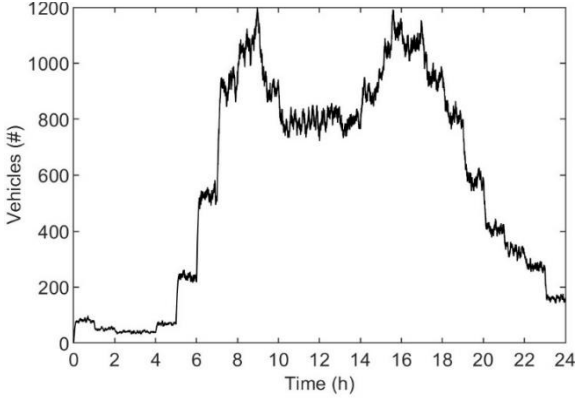

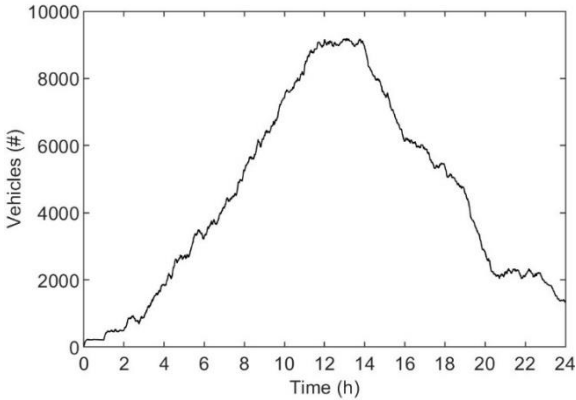

Fig. 2. Traffic demand: (a) real-world profile, (b) artificially inflated profile.

Conservative, moderate, and aggressive behaviors of conventional and automated vehicles were characterized by different values of the `lcAssertive` parameter [14], mimicking the heterogeneity in human driver behavior and automated driving system characteristics. In each simulation run and scenario, there was one third conservative, one third moderate, and the remaining third aggressive vehicles.

*C. Demand modeling and routing*

In the present work, two 24-hour demand profiles were considered. The first profile (Fig. 2a) was derived from actual traffic counts provided by the City of Hannover Traffic Management Center; the second (Fig. 2b), artificially created to observe the network in congested conditions, was obtained by gradually increasing demand up to a saturation point. The former was used for investigating how the number of lane-changing maneuvers, travel time, and average speed was affected by AV penetration rate (Sections IV.A and IV.B). In contrast, the latter was used to evaluate the full spectrum of network performance, which could not be possible with the real-world profile, due to the resulting lack of traffic congestion (Section IV.C).

Regarding the real-world profile, it was representative of the typical working day in the period from January 2018 to February 2020. Therefore, only counts from Tuesday to Thursday were considered and public holidays were excluded. Within the demand profile, an initial peak in demand can be seen within the time interval from 8 to 9 am; the second peak is reached between 4 and 6 pm. Both peaks are due to home-to-work or home-to-school commuting.

For routing, a real-world hourly OD matrix was used to load the demand. The calculated route was the fastest route according to the traffic conditions, calculated using the A* algorithm, which is a graph-based routing and optimization algorithm [27].

*D. Simulation setup*

As the main objective of the present work was to assess the impact of vehicles equipped with automated lane-changing behavior on system efficiency, several scenarios were evaluated. The first was a baseline scenario called "Scenario 0", in which only CVs were loaded onto the network. Ten more scenarios were then analyzed, progressively increasing the penetration rate of AVs by 10% until a 100% AV scenario was reached.

For each of these eleven scenarios, 20 simulation runs were carried out using the real-world traffic demand profile, and the other 25 simulation runs using the artificial demand profile, allowing an evaluation of the network in a realistic situation and also to estimate network performance when reaching capacity. The number of runs was chosen after a sensitivity analysis of vehicles average speed across the runs, following FHWA prescriptions [28].

IV. RESULTS AND DISCUSSION

*A. Number of lane-changing maneuvers*

SUMO lane change output was collected in order to count and identify the lane-changing maneuvers carried out by the vehicles. In the simulation, a continuous lane change was implemented, which means that the vehicle performed a continuous lane-changing maneuver from one lane to the adjacent instead of disappearing from the current lane and reappearing in the adjacent lane.

Fig. 3a reports the percentage change in the total number of lane-changing maneuvers with reference to Scenario 0. A progressive reduction in the number of maneuvers can be observed, as AV penetration rate increased. In the 100% AV scenario, a 26.9 % reduction was reported.

Fig. 3b compares the percentage of AVs in circulation and the percentage of lane-changing maneuvers performed by AVs. The former corresponds to the diagonal of the graph, while the latter follows a trend that is positioned below the diagonal. For example, considering a penetration rate of 50 %, it results that the lane-changing maneuvers carried out by AVs were only 42.8 %, while CVs carry out the remaining 57.2 % of the maneuvers. For each penetration rate, it appears that AVs perform proportionally fewer lane-changing maneuvers than CVs. From these results, it can be deduced that, compared to CVs, AVs follow a more precautionary behavior in lane-changing maneuvers, with the decreasing number of maneuvers performed likely attributable to larger gaps accepted by AVs and, although this might have positive impacts in terms of safety, it could also affect system efficiency.

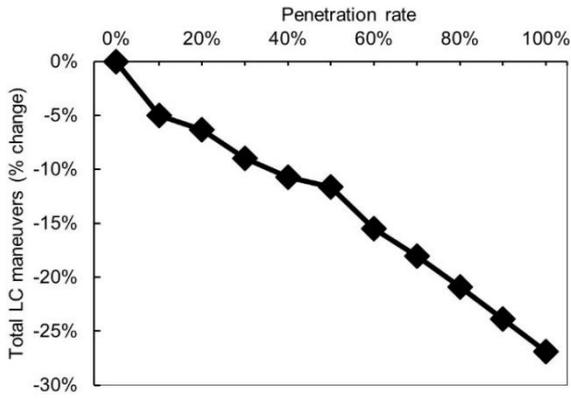

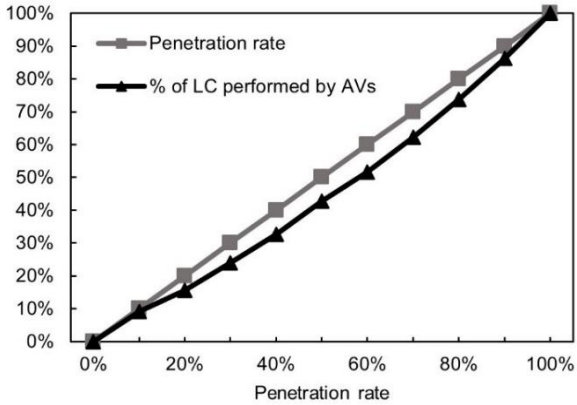

Fig. 3. Number of lane-changing (LC) maneuver: (a) relative change with reference to Scenario 0; (b) comparison between AV penetration rate and the percentage of total lane-changing maneuvers performed by AVs.

### B. Travel time and average speed

SUMO edge-based measures were collected and aggregated to determine the travel time and average speed of the vehicles within the network. In this sense, travel time was defined as the total time spent by vehicles within the network, whereas speed was defined as the average space speed in the network, calculated by weighting the average space speed of a link by the number of vehicles in circulation [7]:

$$\bar{v} = \frac{\sum_{i=1}^{N} v_i N_i}{\sum_{i=1}^{N} N_i} \quad (1)$$

Where $\bar{v}$ is the mean space speed in the network, $i=1,…,N$ is the generic link in the network, $v_i$ is the mean space speed in $i$, and $N_i$ is the number of circulating vehicles in $i$.

Fig. 4 illustrates the results, presented in relative terms to Scenario 0. With regard to the mean travel time, it can be observed that in each scenario considered, AVs showed values that were always higher than CVs. Furthermore, as AV penetration rate increased, also the mean travel time increased, up to 7.35% in the 100% AV scenario. Consistent with this, in each scenario considered, AVs traveled with an average speed lower than those of the CVs. There was also a decrease in average travel speed as the penetration rate of AVs increased, up to a decrease of 6.17 % in the 100% AV scenario.

These descriptive results were confirmed by a statistical analysis carried out with linear mixed models, considering as fixed effect factors the penetration rate and vehicle type (conventional vs. automated), and as random effect grouping factor the simulation seed. In particular, there was a significant effect of penetration rate, $F(1,377)=3127.3$, $p<.001$, and of vehicle type, $F(1,377)=7.1$, $p=.008$, on travel time, and also a significant effect of penetration rate, $F(1,377)=5053.0$, $p<.001$, and of the vehicle type, $F(1,377)=12.1$, $p<.001$, on average speed.

The results indicate a progressive decrease in system efficiency, as the penetration rate increases, with AVs more penalized than CVs. Again, this is likely imputable to the more prudent approach of AVs to the lane-changing maneuver, which causes them to wait longer before being able to perform the desired maneuver. It should be noted, however, that while this decrease is statistically significant, it can be considered quite modest in absolute terms: e.g., in the 100% AV scenario, the travel time increases by about 21 seconds compared to the baseline.

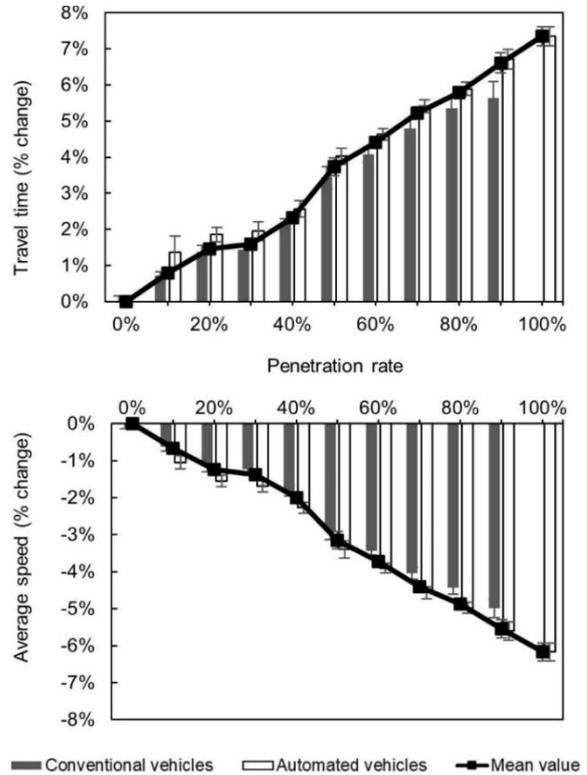

Fig. 4. Relative change of travel time and average speed with reference to Scenario 0.

## C. Network performance

To further investigate overall network performance, macroscopic fundamental diagrams (MFD) were built. Contrary to the analysis presented in Sections IV.A and IV.B, which used a real-world demand profile, here an artificially inflated demand profile was loaded in the network (see Section III.C): this allowed the investigation of the network at the critical point, which could not be reached using the real-world profile.

Under the assumption of stationary flow conditions, the aggregation of edge variables into network variables was determined with the following equations [7], which are consistent with the general MFD formulation proposed by Daganzo and Geroliminis [29]:

$$K = \frac{\sum_{i=1}^{N} N_i}{\sum_{i=1}^{N} l_i} \quad (2)$$

$$Q = \sum_{i=1}^{N} v_i k_i \quad (3)$$

Where K and Q are respectively network density and flow, $i$, $v_i$, $N_i$ are the same as in (1), $l_i$ is the length of $i$, $k_i$ the density of $i$.

Flow and density values were computed for each 15-minute interval of each simulation run of each penetration rate investigated. Fig. 5 presents the flow-density MFD of all simulation seeds for the 50% AV scenario; its shape is qualitatively similar to that of the other scenarios. The diagram shows a considerable dispersion of values, especially in the congested part; the presence of two clusters of points in the congested part of the diagram can also be observed, and it is caused by a hysteresis loop, with the upper part of the diagram relating to flow-density values during network loading and the lower part during network unloading.

The critical point of the diagram was used to investigate how the overall network performance changes with penetration rate. For each scenario, the maximum flow $Q_{max}$ values and optimum density $K_0$ values were obtained by averaging the values of each individual run (Table II). The maximum flow measurements obtained are somewhat consistent with those available in the literature for cities and contexts similar to the city of Hannover [30].

To evaluate whether the penetration rate had a significant effect on the investigated variables, a statistical analysis with linear mixed models was carried out, considering as fixed effects factor the penetration rate, and as random effects grouping factor the simulation seed. A significant effect of the penetration rate was observed on $Q_{max}$, $F(1,249)=53.5$, $p<.001$, but not on $K_0$, $F(1,249)=1.5$, $p=.225$. $Q_{max}$ decreased from 392.98 veh/h/lane in Scenario 0 to 378.62 veh/h/lane in the 100% AV scenario, a 3.65% reduction.

These results are consistent with the analysis carried out in Sections IV.A and IV.B, and indicate a modest, yet statistically significant, decrease in the system efficiency at the critical point.

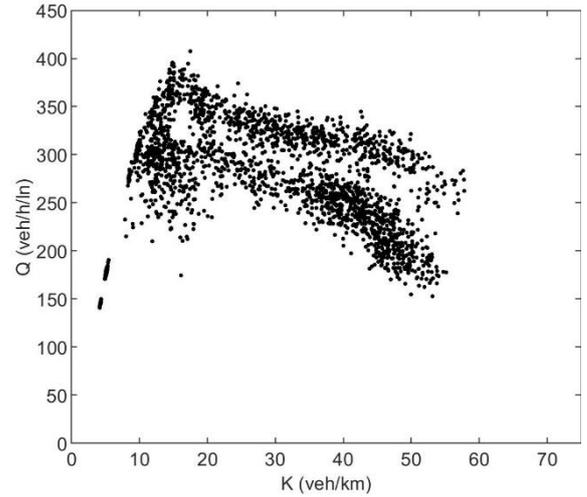

Fig. 5. Flow-density MFD for the 50% AV scenario.

TABLE II. MAXIMUMUM FLOW AND OPTIMUM DENSITY AS AV PENETRATION RATE INCREASES. REPORTED VALUES ARE AVERAGED ACROSS THE 25 SIMULATION SEEDS, STANDARD DEVIATION IN PARENTHESIS

| AV penetration rate | $Q_{max}$ [veh/h/lane] | $K_0$ [veh/km] |
|---|---|---|
| 0% | 392.98 (11.78) | 15.96 (1.44) |
| 10% | 390.83 (12.28) | 15.63 (1.36) |
| 20% | 385.55 (15.27) | 16.00 (1.38) |
| 30% | 387.34 (9.00) | 15.93 (1.27) |
| 40% | 388.98 (8.83) | 15.76 (1.19) |
| 50% | 385.22 (8.46) | 15.68 (1.49) |
| 60% | 383.13 (11.42) | 16.10 (1.49) |
| 70% | 379.72 (10.29) | 15.75 (1.39) |
| 80% | 378.86 (8.68) | 15.75 (1.21) |
| 90% | 377.40 (10.77) | 15.33 (1.68) |
| 100% | 378.62 (12.02) | 15.64 (1.76) |

## V. CONCLUSION

The present study investigated the effect of the lane-changing behavior of AVs on urban network efficiency, considering incrementing levels of penetration rate. The study was carried out with microsimulation software SUMO, and the AVs were modeled to have an automatic lateral control, while keeping a conventional longitudinal control, thus isolating the effects of the automatic lane-changing behavior.

According to our findings, as the penetration rate of AVs increased, the total number of lane-changing maneuvers decreased, with AVs proportionally performing fewer maneuvers than CVs. Furthermore, travel time tended to increase (up to 7.35% in the 100% AV scenario) and the average speed of vehicles decreased (-6.19%), with an overall reduction in performance as the penetration rate increased, penalizing AVs more than CVs. When considering the network overall performance in congested situations, the network capacity decreased (-3.65%) as the AV rate increased. The observed performance degradation was overall modest, but statistically significant, and is likely imputable to the fact that AVs followed a more prudent behavior in lane-changing maneuvers, with larger gaps accepted by AVs compared to CVs. The main takeaway from this work is that further attention should be paid to developing and testing automated lane-changing behavior, as it can negatively impact traffic efficiency.

These results are in contrast to several other studies in the literature (see Section II), which reported increases in system performance as the penetration rate increased. It is possible that, since these studies considered AV with both longitudinal and lateral automated control, the increase in performance due to the automated car-following behavior was able to overcompensate a potential decrease due to the automated lane-changing behavior.

The present study has several limitations and should be regarded as a first step toward a more complete understanding of the effects of automated lane-changing behavior on urban transportation systems. In order to do so, several activities should be performed in future research. First, these results will be confirmed and generalized by considering alternative (and potentially more realistic) calibrations of CV and AV parameters, by using different microsimulation software, and by analyzing road networks located in other cities / counties. Furthermore, other aspects, such as impacts on road safety and vehicle emissions, will be investigated, as a modest decrease in system efficiency could actually be acceptable if there are tangible improvements in terms of safety and sustainability.

ACKNOWLEDGEMENTS

This work was funded by the University of Padua (project "HuMap" – ROSS_BIRD2222_01).